\documentclass[wssquare]{ws-procs961x669} % for citations in square brackets (consult your editor before picking up this style)

\usepackage[T1]{fontenc}
\usepackage[latin9]{inputenc}
\usepackage[dvipsnames]{xcolor}
\usepackage{babel}
\usepackage{mathtools}
\usepackage{amsmath}
\usepackage{amssymb,accents}
\usepackage{mathrsfs}
\usepackage{cancel}
\usepackage{comment}
\usepackage[unicode=true,pdfusetitle,
 bookmarks=true,bookmarksnumbered=false,bookmarksopen=false,
 breaklinks=true,allcolors=blue,backref=false,colorlinks=true]
 {hyperref}
\usepackage{tikz-cd}
\usepackage{tikz}
\usetikzlibrary{shapes,arrows}

\begin{document}
\title{A model of polymer gravitational waves: theory and some possible observational consequences}

\author{Angel Garcia-Chung}

\address{Departamento de F\'isica, Universidad Aut\'onoma Metropolitana - Iztapalapa, \\ San Rafael Atlixco 186, Ciudad de M\'exico 09340, M\'exico\\
   E-mail: alechung@xanum.uam.mx}

\author{James B. Mertens}

\address{Department of Physics and McDonnell Center for the Space Sciences,
Washington University, St. Louis, MO 63130, USA\\
	E-mail: jmertens@wustl.edu}

\author{Saeed Rastgoo}

\address{Department of Physics and
	Astronomy, York University~\\
	4700 Keele Street, Toronto, Ontario M3J 1P3 Canada\\
	E-mail: srastgoo@yorku.ca}

\author{Yaser Tavakoli}

\address{Department of Physics,
University of Guilan, Namjoo Blv.,
41335-1914 Rasht, Iran}

\address{School of Astronomy, Institute for Research in Fundamental Sciences (IPM),  P. O. Box 19395-5531, Tehran, Iran\\
	E-mail: yaser.tavakoli@guilan.ac.ir}
	
\author{Paulo Vargas Moniz}

\address{Departamento de Fisica, Centro de Matematica e Aplica\c c\~oes:  CMA-UBI, Universidade da Beira Interior, 6200 Covilh\~a,
Portugal\\
	E-mail: pmoniz@ubi.pt}

\begin{abstract}
We propose a polymer quantization scheme to derive the effective propagation of gravitational waves on a classical 
Friedmann-Lemaitre-Robertson-Walker (FLRW) spacetime. These waves, which may originate from
a high energy source, are a consequence of the dynamics of the gravitational field in a linearized low-energy regime. A novel method of deriving the effective Hamiltonian of the system is applied to overcome the challenge of polymer quantizing a time-dependent Hamiltonian. Using such a Hamiltonian, we derive the effective equations of motion and show that (i) the form of the waves is modified, (ii) the speed of the waves depends on their frequencies, and (iii) quantum effects become more apparent as waves traverse longer distances.
\end{abstract}

\keywords{
%Proceedings of MG16 conference; World Scientific Publishing.
Quantum gravity, gravitational waves
}

\bodymatter

\section{Introduction\label{sec:Intro}}
Recent discovery of gravitational waves (GWs) and the rapid increase in the sensitivity of GWs observatories has opened up a great opportunity in connecting theory and phenomenology with experiment in precision cosmology, black hole physics and quantum gravity among other fields. In particular we are now becoming more hopeful about the observation of signatures of quantum gravity in GWs emitted from black hole mergers and high redshift regions of the cosmos.

There have been numerous studies connecting theories of quantum gravity with potential observations regarding the structure of quantum spacetime. In particular, in Loop Quantum Gravity (LQG) \cite{Thiemann:2007pyv}, there have been studies to understand the consequence of nonpertubative quantization in propagation of Gamma Ray Bursts (GRBs), other matter fields, and GWs on cosmological or black holes spacetimes (for some examples see, Refs. \cite{Ashtekar:1991mz,Varadarajan:2002ht,Freidel:2003pu,Bojowald:2007cd,Ashtekar:2009mb,Hossain:2009vd,Gambini:2009ie,Mielczarek:2010bh,Date:2011bg,Gambini:2011mw,Gambini:2011nx,Sa:2011rm,Hinterleitner:2011rb,Hinterleitner:2012zz,Neville:2013wba,Neville:2013xba,Hoehn:2014qxa,Arzano:2016twc,Tavakoli:2015fvz,Dapor:2012jg, ElizagaNavascues:2016vqw,Bonder:2017ckx,Lewandowski:2017cvz,Hinterleitner:2017ard,Dapor:2020jvc,Tavakoli:2014mra,Calcagni:2020ume,Calcagni:2019kzo,Calcagni:2019ngc,Garcia-Chung:2020zyq} and references within).
    
In this work we consider GWs as effective perturbations propagating on a classical FLRW cosmological spacetime. The effective form of such waves is derived by applying the techniques of polymer quantization \cite{Ashtekar:2002sn,Corichi:2007tf,Morales-Tecotl:2016ijb,Tecotl:2015cya,Flores-Gonzalez:2013zuk} to the classical perturbations. Such a quantization is a representation of the classical algebra on a Hilbert space that is unitarily inequivalent to the usual Schr\"{o}dinger representation. In it, operators are regularized and written in a certain exponential form. In such theories, the infinitesimal generators corresponding to some of the operators do not exist on the Hilbert space. As a consequence, the conjugate variables to those operators only admit finite transformations. Thus, the dynamics of the theory leads to the  discretization of the spectrum of the conjugate operators (for more details and some examples of polymer quantization applied to particles and path integral formulation of black holes, see Refs. \cite{garcia2014polymer, garcia2016polymer, Morales-Tecotl:2016ijb,Tecotl:2015cya,Morales-Tecotl:2018ugi}).  

In the model we present in this paper, the Hamiltonian is time-dependent and directly polymer quantizing it proves to be quite challenging. Hence, we apply a novel method based on the use of extended phase space, to overcome this issue (see Ref. \cite{garcia2017dirac}). The Hamiltonian in the extended phase space is rendered time independent by applying a certain canonical transformation and then polymer quantized using some of techniques developed in the literature \cite{austrich2017instanton, Tecotl:2015cya, Morales-Tecotl:2016dma}. After that, the effective version of this quantum Hamiltonian is made time-dependent again by applying the inverse of the above-mentioned canonical transformation. Finally the system is re-expressed in the usual non-extended phase space. Using this modified Hamiltonian, we derive the effective equations of motion of polymerized GWs and show that i) the form of the waves is modified, ii) the speed of the waves depends on their frequencies, and iii) the quantum effects are amplified by the distance/time the waves travel.     

This paper is organized as follows: in Sec. \ref{sec:GRW-H}, we derive the classical Hamiltonian of perturbations on an FLRW classical background. In Sec. \ref{sec:polymerization}, this time-dependent Hamiltonian is turned into a polymer effective time-dependent Hamiltonian by applying a certain method that is inspired by an approach used to deal with time-dependent harmonic oscillators. Using this Hamiltonian, we then derive the effective equations of motion. In Sec. \ref{sec:pert-nonpert}, we study the behavior of the solutions in both nonperturbative and perturbative regimes and show that quantum gravitational effects induce certain imprints on the waveform, frequency, and speed of GWs. Finally, in Sec. \ref{sec:conclusion} we present our concluding remarks.

\section{Hamiltonian formalism for GWs\label{sec:GRW-H}}

We start with a spacetime manifold $M=\mathbb{T}^{3}\times\mathbb{R}$
with a spatial 3-torus topology\footnote{To avoid a discussion of boundary conditions on fields (generated
by perturbations), we will assume that the spatial 3-manifold is $\mathbb{T}^{3}$.}, equipped with coordinates $x^{j}\in(0,\ell)$ and a temporal coordinate
$x^{0}\in\mathbb{R}$. The background metric $\mathring{g}_{\mu\nu}$
is then perturbed by a small perturbation $h_{\mu\nu}$ such that
the full metric $g_{\mu\nu}$ can be written as 
\begin{equation}
g_{\mu\nu}=\mathring{g}_{\mu\nu}+h_{\mu\nu}.\label{Eq:metric-pert}
\end{equation}
GWs are the result of the weak-field approximation to the Einstein
field equations for the above metric. As is well-known, a wave traveling
along, say, the $x^{3}$ direction, can be separated into two polarization
scalar modes $h_{+}(x)$ and $h_{\times}(x)$ as 
\begin{equation}
h_{ij}(x)\,=\,h_{+}(x)e_{ij}^{+}+h_{\times}(x)e_{ij}^{\times}\,,\label{polarizedmetric}
\end{equation}
where 
\begin{align}
e^{+}=\left(\begin{array}{cc}
1 & 0\\
0 & -1
\end{array}\right)\quad\quad\text{and}\quad\quad e^{\times}=\left(\begin{array}{cc}
0 & 1\\
1 & 0
\end{array}\right).
\end{align}
We would like to study the dynamics of the perturbations on a homogeneous,
isotropic universe described by the FLRW metric 
\begin{equation}
\mathring{g}_{\mu\nu}dx^{\mu}dx^{\nu}=-N^{2}(x^{0})\,d(x^{0})^{2}+a^{2}(x^{0})d\mathbf{x}^{2},\label{metric0}
\end{equation}
where $x^{0}$ is an arbitrary time coordinate, $N(x^{0})$ is the
lapse function which depends on the choice of $x^{0}$, and $d\mathbf{x}^{2}=\sum_{i}^{3}d(x^{i})^{2}$
is a unit 3-sphere. To obtain a Hamiltonian which resembles the Hamiltonian
of a harmonic oscillator, we introduce a new field
\begin{equation}
h_{\sigma}\left(x^{0},\mathbf{x}\right)=\frac{a^{2}\sqrt{\kappa}}{\ell^{3/2}}\sum_{\mathbf{k}}\mathcal{A}_{\sigma,\mathbf{k}}\left(x^{0}\right)e^{i\mathbf{k}\cdot\mathbf{x}}
\end{equation}
which together with its conjugate momentum ${\cal E}_{\mathbf{\sigma,k}}$
constitute a canonically conjugate pair
\begin{equation}
\left\{ {\cal A}_{\lambda,\mathbf{k}},{\cal E}_{\mathbf{\lambda^{\prime},k}^{\prime}}\right\} =\delta_{\mathbf{k}\mathbf{k}^{\prime}}\delta_{\lambda\lambda^{\prime}}.\label{eq:PB-AE}
\end{equation}
Here $\ell$ is the result of our quantization on a lattice which
corresponds to an upper limit on the momenta involved. The above canonical
pair allow us to write the Hamiltonian of the system as 
\begin{equation}
H=\frac{N}{2a^{3}}\sum_{\sigma=+,\times}\sum_{\mathbf{k}}\left[{\cal E}_{\mathbf{\sigma,k}}^{2}+k^{2}a^{4}{\cal A}_{\sigma,\mathbf{k}}^{2}\right]\eqqcolon\sum_{\sigma=+,\times}\sum_{\mathbf{k}}H_{\sigma,\mathbf{k}}\left(x^{0}\right),\label{eq:Hamiltonian-FLRW-1}
\end{equation}
where $N$ is the lapse. It is clear that this last equation represents
the Hamiltonian of a set of decoupled harmonic oscillators with a
time-dependent frequency $\omega^{2}=k^{2}a^{4}$, due to time dependence
of $a(t)$.

At this point, we choose the harmonic time gauge where $N(x^{0}=\tau)=a^{3}(\tau)$
to get rid of the factor $a^{-3}$ in front of Eq. \eqref{eq:Hamiltonian-FLRW-1}.
Hence, the Hamiltonian of the perturbations (for a fixed mode $\mathbf{k}$
and a fixed polarization $\sigma$) over the FLRW background in harmonic
time becomes 
\begin{equation}
H_{\sigma,\mathbf{k}}(\tau)=\frac{1}{2}\left[{\cal E}_{\mathbf{\sigma,k}}^{2}+k^{2}a^{4}{\cal A}_{\sigma,\mathbf{k}}^{2}\right].\label{eq:Hamiltonian-FLRW-2}
\end{equation}

\section{Polymer quantization and the effective Hamiltonian\label{sec:polymerization}}

The time-dependence of this Hamiltonian \eqref{eq:Hamiltonian-FLRW-2}
makes deriving its effective polymer corrections quite complicated.
This is because its polymer quantization will yield a time-dependent
quantum pendulum-type system whose solutions are mathematically difficult
to treat. In order to circumvent this issue, we will apply a procedure
based on the extended phase space formalism (more details in Ref.
\cite{garcia2017dirac}). The idea is to first lift the (action of
the) system 
\begin{equation}
S=\int\left\{ p\frac{dq}{dt}-H(t)\right\} dt,\label{TDHO-1}
\end{equation}
with time-dependent Hamiltonian of the form 
\begin{equation}
H(t)=\frac{1}{2m}p^{2}+\frac{1}{2}m\omega(t)^{2}q^{2},
\end{equation}

\noindent to the extended phase space (EPS). In accordance with Dirac's
formalism, the system is now described by the extended action 
\begin{equation}
S=\int\left\{ p\frac{dq}{d\tau}+p_{t}\frac{dt}{d\tau}-\lambda\phi\right\} d\tau,\label{TDHOExt-1}
\end{equation}
where 
\begin{equation}
\phi=p_{t}+H(t)\approx0,
\end{equation}
is a first class constraint ensuring the compatibility of the two
actions \eqref{TDHO-1} and \eqref{TDHOExt-1} on the constrained
surface $\phi=0$, $\lambda$ is a Lagrange multiplier, and $p_{t}$
is the momentum conjugate to $t$ .This is step (1) in Fig. \ref{fig:schem}.
In step (2) in Fig. \ref{fig:schem}, we apply a canonical transformation
\begin{align}
Q & =\frac{1}{\rho(t)}q,\label{eq:can-tr-1}\\
T & =\int\frac{1}{\rho^{2}(t)}\,dt,\label{eq:can-tr-2}\\
P & =\rho(t)p-m\dot{\rho}(t)q,\label{eq:can-tr-3}\\
P_{T} & =\rho^{2}(t)p_{t}+\rho(t)\dot{\rho}(t)\,q\,p-\frac{m}{2}q^{2}\left[\dot{\rho}^{2}(t)+\frac{W^{2}}{\rho^{2}(t)}-\omega^{2}(t)\rho^{2}(t)\right],\label{eq:can-tr-4}
\end{align}
in the extended phase space which removes the time dependency of the
Hamiltonian $H(t)$ in $\phi$. Here, $W$ is the time-independent
frequency of the time-independent system and $\rho$ is an auxiliary
variable to be determined by the specific properties of the system,
more precisely by $\omega$ and $W$. After this step the action becomes

\begin{equation}
S=\int\left\{ P\frac{dQ}{d\tau}+P_{T}\frac{dT}{d\tau}-\lambda\tilde{\phi}\right\} d\tau,\label{TDHONew-1}
\end{equation}
where, the first class constraint now reads 
\begin{equation}
\tilde{\phi}=\rho^{2}(T)\left[P_{T}+K\right]\approx0,\label{eq:phi-tild-class}
\end{equation}
and the corresponding Hamiltonian $K$ appearing in it is 
\begin{equation}
K=\frac{1}{2m}P^{2}+\frac{1}{2}mW^{2}Q^{2}.\label{NewHo-1}
\end{equation}

\noindent Moreover, the auxiliary equation used to fix $\rho(t)$
becomes 
\begin{equation}
\ddot{\rho}(t)+\omega^{2}(t)\rho(t)=\frac{W^{2}}{\rho^{3}(t)}.\label{eq:rho-eom}
\end{equation}
This time-independent harmonic oscillator can now be polymer quantized
\cite{austrich2017instanton,Tecotl:2015cya,Morales-Tecotl:2016dma}
as in step (3) of Fig. \ref{fig:schem}. We then perform the inverse
canonical transformations above in step (4) of Fig. \ref{fig:schem}.
Finally, in step (5) of Fig. \ref{fig:schem}, we apply the inverse
of the canonical transformations above and solve the constraint. This
yields the polymer effective Hamiltonian on the usual phase space,
where now the Hamiltonian is not just effective but also time-dependent.

%%%%%%%%%%%%%%%
\tikzstyle{block} = [rectangle, draw, fill=blue!20, 
text width=7.5em, text centered, rounded corners, minimum height=3em]
\tikzstyle{line} = [draw, -latex']
%%%%%%%%%%%%%%%%
\begin{figure}[!t]
  \centering
  %\hspace*{-100pt}
  {\scriptsize
  \begin{tikzpicture}[node distance = 2cm, auto]
      
    \node [block] (consent) { $(q(t),p(t)); H(t)$};
    
    \node [block, above of = consent, node distance = 2.0cm,text width=5.8em] (screening) {$(q,p,t,p_t)$;\\ $\phi=p_t-H(t)$};
    
    \node [block, right of = screening, node distance = 4.1cm,text width=11.6em] (refer) {$(Q,P,T,P_T)$;\\ $\tilde{\phi}=\rho^{2}(T)\left[P_{T}-K(Q,P)\right]$};
    
    \node [block, right of = refer, node distance = 3.55cm,text width=4.4em] (refer2) { $K_{\textrm{eff}}(Q,P)$};
    
    \node [block, right of = refer2, node distance = 2.55cm,text width=6.1em] (refer3) { $K_{\textrm{eff}}(q,p,t,p_t)$};
    
   \node [block, below of = refer3, node distance = 2.0cm] (refer4) { $K_{\textrm{eff}}(q,p)$};
  
    % edges
    \path [line] (consent) -- node {\textrm{(1) to EPS}}(screening);
    \path [line] (refer) -- node { \textrm{(3) Poly.}}(refer2); 
    \path [line] (refer2) -- node { \textrm{(4) CT}}(refer3);
    \path [line] (screening) -- node { \textrm{(2) inv. CT}}(refer);
    \path [line] (refer3) -- node {(5) to PS}(refer4);
     
  \end{tikzpicture}}
  \caption{Schematics of the derivation of a time-dependent effective Hamiltonian constraint. Here ``EPS'' means extended phase space, ``inv. CT'' denotes inverse canonical transformation, ``Poly.'' means the process of polymer quantization and getting an effective polymer Hamiltonian from there, ``CT'' denotes the canonical transformation, and ``PS'' means the nonextended phase space. The lower row corresponds to the usual phase space, while the upper row corresponds to the extended phase space. \label{fig:schem}}
\end{figure}
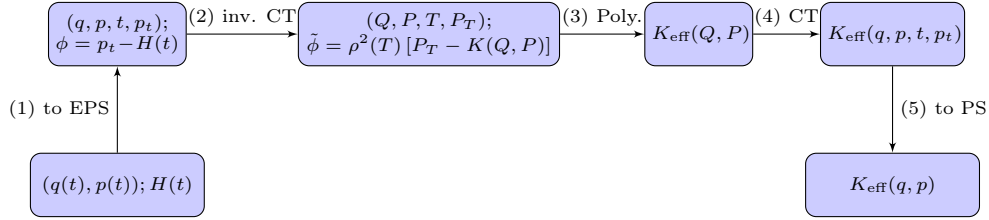

%%%%%%%%%%%%%%%%%%%%%%%%%%%%%%%%%%%%%%%%%%%%%%%%%
By applying this method to our Hamiltonian \eqref{eq:Hamiltonian-FLRW-1},
we obtain an effective polymer Hamiltonian with polymer $\mathcal{E}_{\sigma,\mathbf{k}}$
and discrete spectra for $\mathcal{A}_{\sigma,\mathbf{k}}$ as

\begin{align}
H_{\textrm{eff}}^{(\mathcal{E})}=\sum_{\sigma=+,\times}\sum_{\mathbf{k}\in\mathscr{L}} & \left\{ \frac{2}{\mu^{2}\rho^{2}}\sin^{2}\left(\frac{\mu\left(\rho\mathcal{E}_{\sigma,\mathbf{k}}-\dot{\rho}\mathcal{A}_{\sigma,\mathbf{k}}\right)}{2}\right)+\frac{\dot{\rho}\mathcal{A}_{\sigma,\mathbf{k}}\,\mathcal{E}_{\sigma,\mathbf{k}}}{\rho}+\frac{\mathcal{A}_{\sigma,\mathbf{k}}^{2}}{2}\left[\omega^{2}-\frac{\dot{\rho}^{2}}{\rho^{2}}\right]\right\} ,\label{eq:H-E-eff}
\end{align}
where $\mu$ is called the polymer parameter that sets the scale for
which the quantum gravity effects become important, and we have set
$\hbar=1$. The corresponding equations of motion read 
\begin{align}
\frac{d{\cal A}_{\sigma,\mathbf{k}}}{dt} & =\frac{1}{\rho}\frac{\sin\left(\mu\left(\rho\mathcal{E}_{\sigma,\mathbf{k}}-\dot{\rho}\mathcal{A}_{\sigma,\mathbf{k}}\right)\right)}{\mu}+\frac{\dot{\rho}}{\rho}\mathcal{A}_{\sigma,\mathbf{k}},\label{eq:EoM-eff-E-1}\\
\frac{d{\cal E}_{\sigma,\mathbf{k}}}{dt} & =\frac{\dot{\rho}}{\rho^{2}}\frac{\sin\left(\mu\left(\rho\mathcal{E}_{\sigma,\mathbf{k}}-\dot{\rho}\mathcal{A}_{\sigma,\mathbf{k}}\right)\right)}{\mu}+\left(\frac{\dot{\rho}}{\rho}\right)^{2}\mathcal{A}_{\sigma,\mathbf{k}}-\omega^{2}\mathcal{A}_{\sigma,\mathbf{k}}-\frac{\dot{\rho}}{\rho}\mathcal{E}_{\sigma,\mathbf{k}}.\label{eq:EoM-eff-E-2}
\end{align}
These equations are nonlinear in both ${\cal A}_{\sigma,\mathbf{k}}$
and $\mathcal{E}_{\sigma,\mathbf{k}}$, and their $\mu\to0$ limit
matches the classical equations of motion as expected. Also notice
that in our case, $\rho$ controls the background geometry such that
a time-dependent $\rho$ corresponds to a time-dependent background
geometry and a constant $\rho$ corresponds to the flat spacetime.

\section{Perturbative and nonperturbative numerical solutions\label{sec:pert-nonpert}}

We will now solve Eqs.~\eqref{eq:EoM-eff-E-1}--\eqref{eq:EoM-eff-E-2}
for specific field-space configurations, both perturbatively, and
numerically and nonperturbatively for both time-dependent and time-independent
backgrounds.

\subsection{Time-independent background}

\begin{figure}[tb]
\includegraphics[width=1\textwidth]{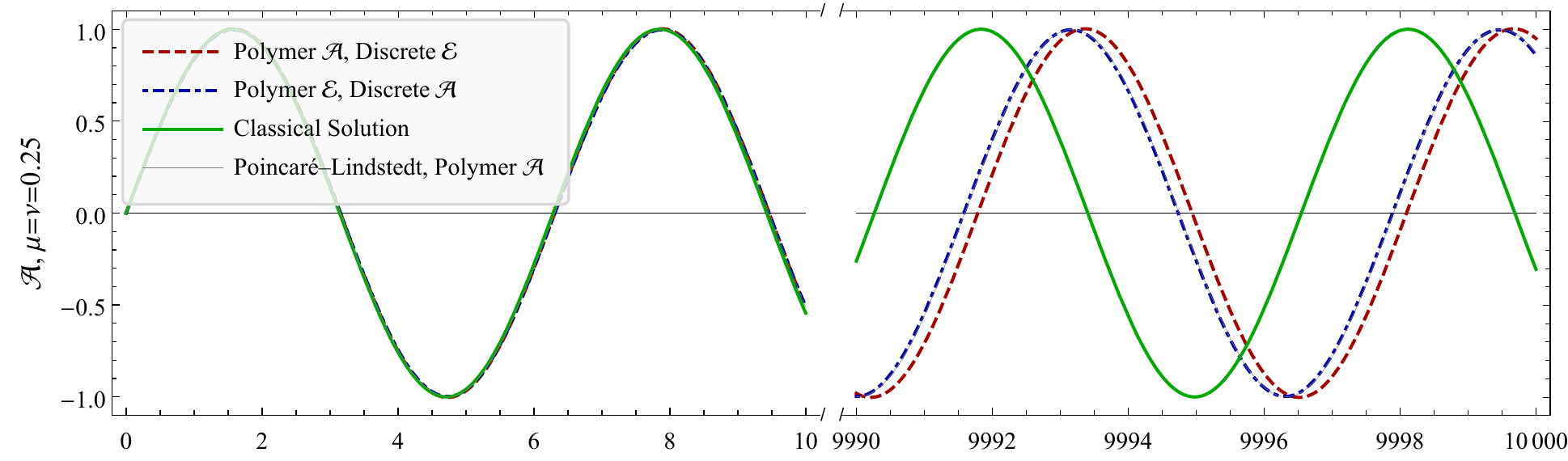} \includegraphics[width=1\textwidth]{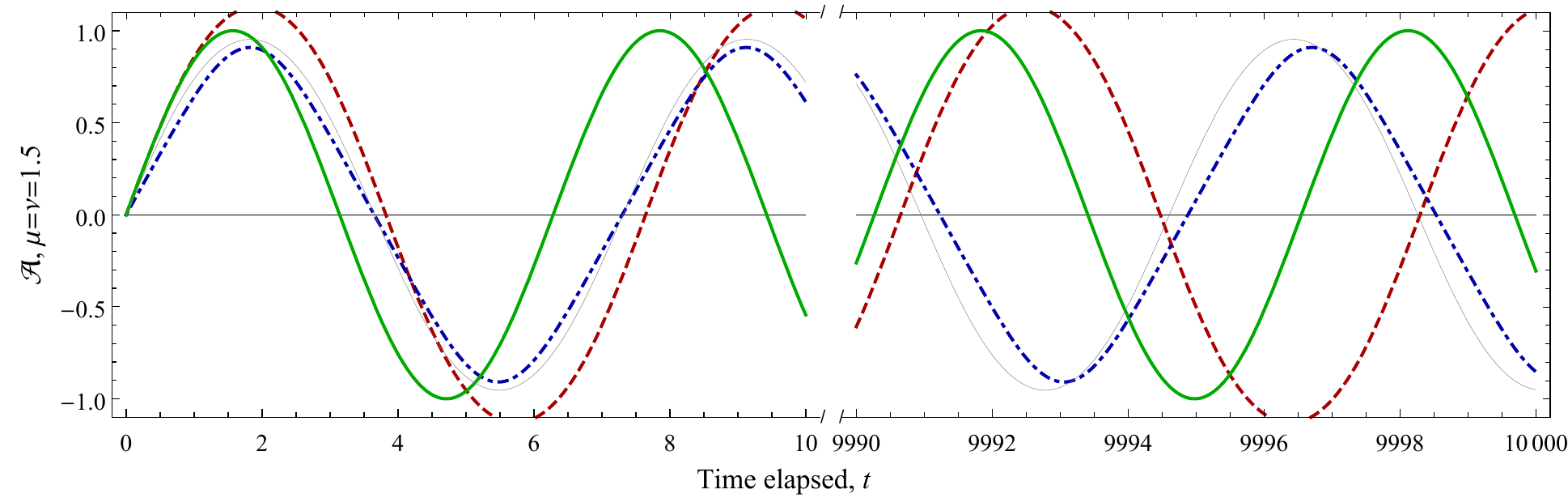}
\caption{ Time evolution of $\mathcal{A}$ with $\mathcal{A}_{I}=0$, $\mathcal{E}_{I}=1$,
and $k=1$ for two different choices of $\mu$ in the case of
a time-independent background spacetime, i.e., $\rho=$\, const. Here $\nu$ refers to another representation of the polymer quantization in which the momentum $\mathcal{E}$ is discrete, which can be found in our original paper \citep{Garcia-Chung:2020zyq}. 
The solutions are shown at early times, and the axis is broken to
show the behavior at a much later time. Solutions can be mapped to
different choices of $k$ and $\mathcal{E}_{I}$, while
changing $\mathcal{A}_{I}=0$ can be viewed as a phase shift. \label{fig:A-time-evol}}
\end{figure}

For a time-independent background, for which $\rho=1$ and $\dot{\rho}=\ddot{\rho}=0$,
we can obtain full nonpertubative numerical solutions for $\mathcal{A}(t)$
as seen in Fig.~\ref{fig:A-time-evol}. We can also obtain a perturbative
solution
\begin{align}
\mathcal{A}(t)\simeq & \,\,\mathcal{E}_{I}\sin\left[(1-(\mathcal{E}_{I}k\mu)^{2}/16)kt\right]\nonumber \\
 & -\frac{\mathcal{E}_{I}^{3}k^{2}\mu^{2}}{16}\sin^{2}\left[(1-(\mathcal{E}_{I}k\mu)^{2}/16)kt\right]\cos\left[(1-(\mathcal{E}_{I}k\mu)^{2}/16)kt\right]\,.\label{eq:polyE-Abar-pert}
\end{align}
This solution can exhibits a frequency shift of order $\mu^{2}$,
and a cubic correction term. The cubic term can also be rewritten,
and thought of, as an introduction of higher harmonics using angle
identities. In observations, the frequency shift may be more important
to account for than the excited harmonics. This is because the frequency
shift can manifest as a phase shift that has considerable time to
develop as the wave traverses cosmological distances. In Fig.~\ref{fig:A-time-evol}
we demonstrate this, comparing the perturbative solution to the exact
and classical ones for the time-independent case.

We can also analyze the above perturbative solutions and obtain some
insight into the speed of propagation of the waves. For that, we note
that the dominant contributions to Eq. \eqref{eq:polyE-Abar-pert}
can be written as 
\begin{equation}
\mathcal{A}(t)\simeq\mathcal{E}_{I}\sin\left[\left(1-\left(\frac{\mathcal{E}_{I}k\mu}{4}\right)^{2}\right)kt\right].
\end{equation}
Comparing with the classical solution where we identify $ka^{2}=\omega_{c}$,
with $\omega_{c}$ being the classical angular speed, we notice that
up to first order the polymer angular speed is 
\begin{equation}
\omega_{\mu}^{(\mathcal{E})}\,\simeq\,\omega_{c}\left[1-k^{2}\left(\frac{\mathcal{E}_{I}\mu}{4}\right)^{2}\right].\label{eq:omega-mu-approx}
\end{equation}
Although this is a perturbative and approximate result and even though
we have neglected higher harmonics in Eq. \eqref{eq:polyE-Abar-pert},
the above equation reveal a curious phenomenon. Noting that $\omega_{c}=ka^{2}$
and with the group velocity being 
\begin{equation}
v=\frac{d\omega_{\textrm{poly}}}{d\left(ka^{2}\right)}
\end{equation}
with $\omega_{\textrm{poly}}$ being either $\omega_{\nu}^{(\mathcal{A})}$
or $\omega_{\nu}^{(\mathcal{E})}$, we obtain 
\begin{equation}
v_{\mu}^{(\mathcal{E})}\,\simeq\,1-k^{2}\left(\frac{\mathcal{E}_{I}\mu}{4}\right)^{2}.\label{eq:grp-v-E}
\end{equation}
where $v_{\mu}^{(\mathcal{E})}$ is the velocity of the effective
waves. One can see from Eq. \eqref{eq:grp-v-E} that the group velocity
of the waves is slower than the speed of light by a factor of $k^{2}\left(\frac{\mathcal{E}_{I}\mu}{4}\right)^{2}$,
and it also depends on the initial momentum $\mathcal{E}_{I}$ of
the waves and the polymer parameter $\mu$ due to the factor $k^{2}\left(\frac{\mathcal{E}_{I}\mu}{4}\right)^{2}$.
More importantly, the deviation from the speed of light also depends
on the modes $k$. Hence, waves with larger $k$ (i.e., larger energies)
have a lower speed compared to the ones with smaller $k$ and are
more affected by the quantum structure of spacetime. Also, notice
that this case leads to the violation of Lorentz symmetry as can be
seen by squaring both sides of Eq. \eqref{eq:omega-mu-approx}. Of
course, due to the sheer smallness of the expected value of $\mu$,
and the appearance of their squares in the above expressions, these
effects are very small, but a highly energetic phenomenon with a large
$\mathcal{E}_{I}$ may help to amplify it to an extent that future
observatories can detect it. We should emphasize that the presence
of the violation of the Lorentz symmetry in this case, as seen from
the above results, is a consequence of the polymer quantization and,
in particular, this model, and is not a direct consequence of LQG.

\subsection{Time-independent background}

\begin{figure}[tb]
\centering \includegraphics[width=0.6\textwidth]{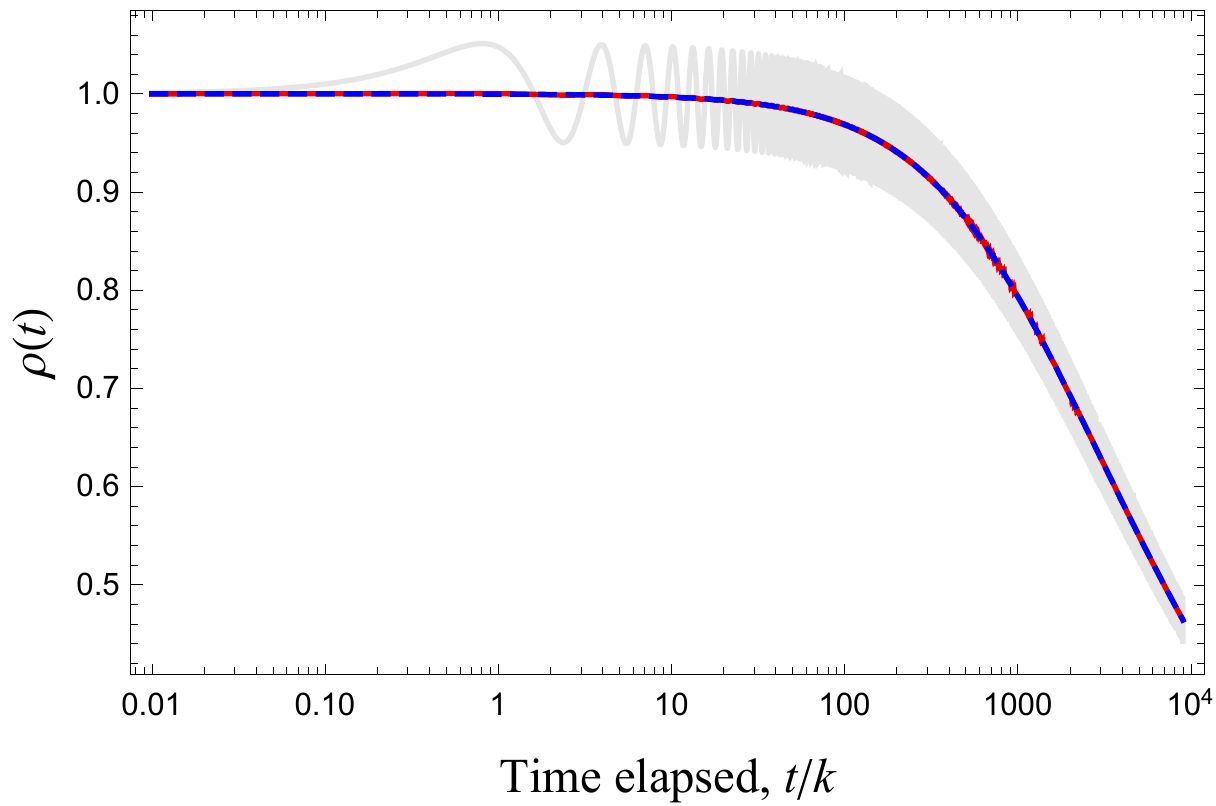} \caption{Evolution of the auxiliary variable $\rho(t)$. The full numerical
nonoscillatory solution is shown in solid red, an approximate power-law
solution is shown in dashed blue, and a solution with initial conditions
that result in oscillatory behavior is shown in light grey.}
\label{fig:rho-t} 
\end{figure}

\begin{figure}[tb]
\includegraphics[width=1\textwidth]{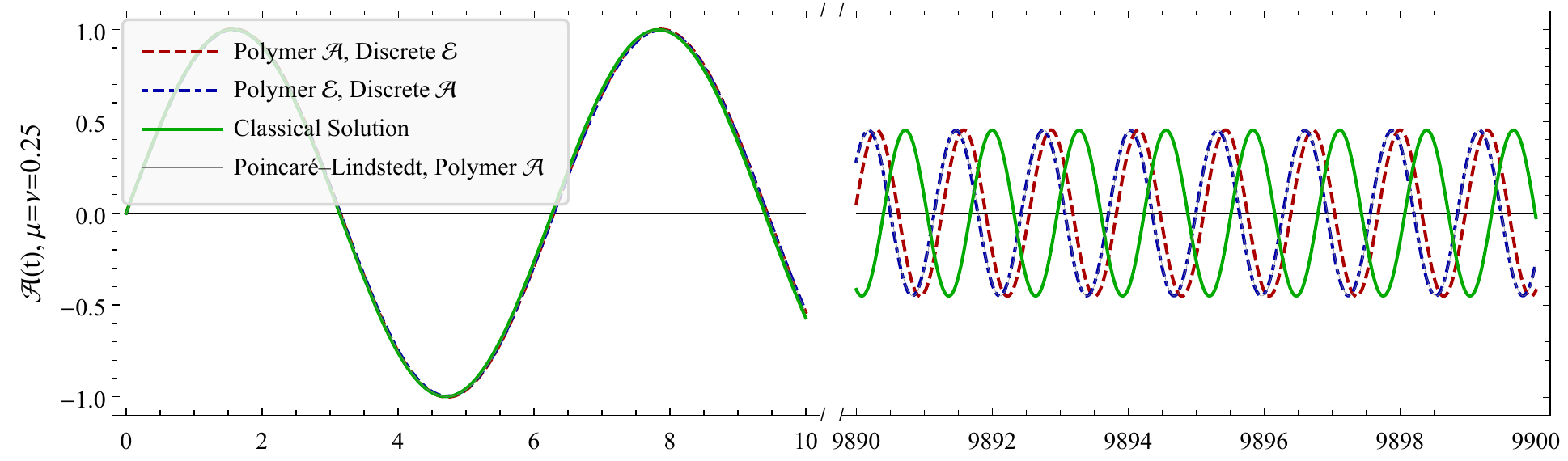} \includegraphics[width=1\textwidth]{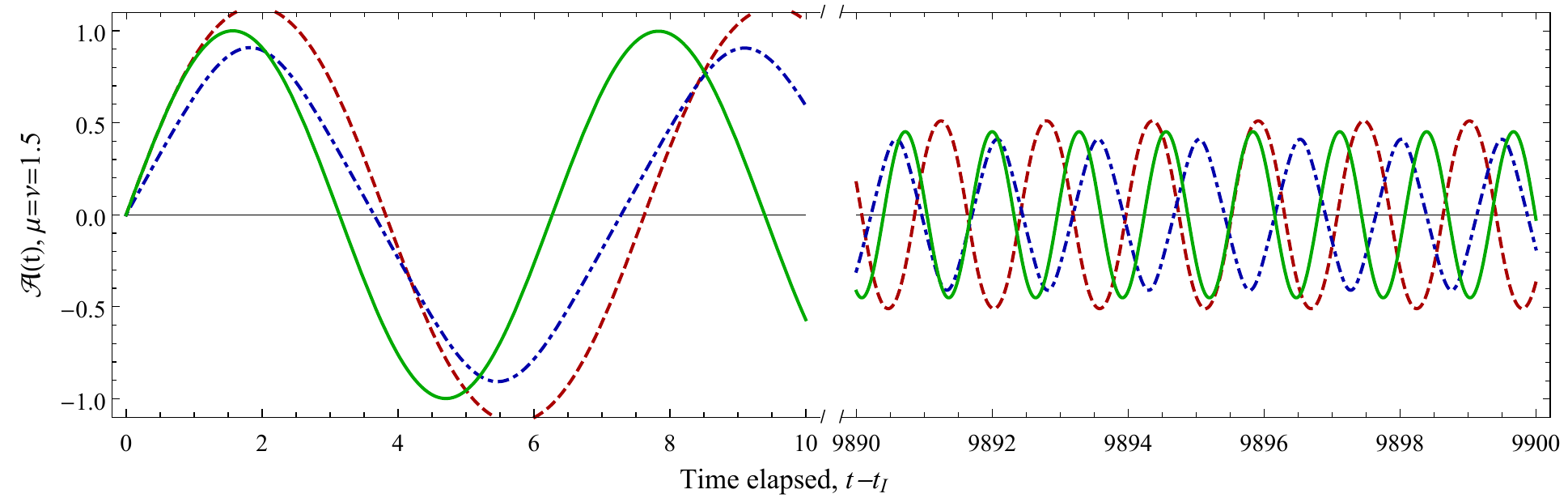}
\caption{Time evolution of $\mathcal{A}$ (as in Fig.~\ref{fig:A-time-evol})
for two different choices of $\mu$, for the case of a time-dependent
background, i.e., $\rho(t)$ as described in the text. The axis is
broken to show the behavior at a later time. Again, $\nu$ refers to another representation of the polymer quantization in which the momentum $\mathcal{E}$ is discrete, which can be found in our original paper \citep{Garcia-Chung:2020zyq}. \label{fig:A-time-evol-flrw}}
\end{figure}

For the case of a time-dependent background, we can obtain a solution
in one of two ways: directly integrating the EOMs, or using the canonical
transformation in Eqs.~\eqref{eq:can-tr-1}--\eqref{eq:can-tr-4}.
In either case, we will need to obtain a solution for $\rho$ by solving
Eq.~\eqref{eq:rho-eom}. In general, this choice determines whether
the mode amplitude will be purely decaying or will contain oscillatory
behavior. Here we will seek purely growing solutions for $\rho$,
choosing initial conditions such that oscillatory behavior is minimized;
in our case, simply choosing $\rho=1$ and $\dot{\rho}=0$ is sufficient.
Choosing a different initial amplitude for $\rho$ is in any case
equivalent to rescaling of the scale factor $a$, polymer scale, momentum,
and time coordinate. The full nonperturbative solutions are plotted
in Fig.~\ref{fig:A-time-evol-flrw}. We can also obtain a perturbative
solution 
\begin{align}
\mathcal{A}(t)\simeq & \,\,\mathcal{E}_{I}\rho\sin\left[(1-(\mathcal{E}_{I}k\mu)^{2}/16)kT(t)\right]\nonumber \\
 & -\frac{\mathcal{E}_{I}^{3}k^{2}\mu^{2}}{16}\rho\sin^{2}\left[(1-(\mathcal{E}_{I}k\mu)^{2}/16)kT(t)\right]\cos\left[(1-(\mathcal{E}_{I}k\mu)^{2}/16)kT(t)\right]\,,\label{eq:polyE-Abar-pert2}
\end{align}
where 
\begin{equation}
T(t)=\int_{t_{I}}^{t}dt'\frac{1}{\rho(t')^{2}}
\end{equation}
For GWs emitted at a time much greater than the characteristic wave
time scale, i.e., $t_{I}\gg k^{-1}$, where $T_{I}$ is the initial
time, and for nonoscillatory solutions of $\rho$, the second-derivative
term is small, and solutions to the auxiliary equations are well approximated
by a simple power law, $\rho=1/a$. In Fig.~\ref{fig:rho-t} we show
the behavior of $\rho$ for several sets of initial conditions, and
for a universe with a cosmological constant with $w=-1$, $a\propto t^{1/3}$,
and $t_{I}=10^{3}$ (in units of $k^{-1}$). In subsequent plots we
will use initial conditions that do not result in oscillatory behavior.

From the canonical transformation~\eqref{eq:can-tr-1}--\eqref{eq:can-tr-3}
(or, rather, its inverse), we see that the time-dependent waveform
amplitude will pick up an overall factor of $\rho$ relative to the
time-independent one, the time coordinate will be altered, and the
momentum will be similarly rescaled but will also pick up an additional
factor proportional to the wave amplitude. Due to the monotonically
decreasing nature of $\rho$ and the smallness of its derivative,
this additional factor will be a strongly subdominant contribution.
In Fig.~\ref{fig:A-time-evol-flrw} we show the final solution for
the field $\mathcal{A}(t)$ for this time-dependent background. Somewhat
counterintuitively, the frequency is seen to increase at later times;
more commonly the frequency is considered to decrease (redshift) with
cosmological expansion. This is due to the choice of harmonic slicing
we have made, with $N=a^{3}$ instead of the more commonly used $N=1$
(synchronous) or $N=a$ (comoving) time coordinate.

\section{Discussion and Conclusion\label{sec:conclusion}}

In this work we have studied a certain effective form of GWs, considered
as quantized perturbations propagating over a classical FLRW spacetime,
in order to derive observational signatures to be compared with the
results of experiments conducted by GW observatories. We have considered
the Hamiltonian of classical gravitational perturbations, a time-dependent
Hamiltonian, and have applied the techniques of polymer quantization
to it. This polymer quantization was applied to each of the Fourier
modes of the GW. A feature of this quantization is that the one-particle
Hilbert space is modified and the Lorentz symmetry is no longer present
\cite{garcia2016polymer}. This modification is ``encoded'' on the
polymer scale $\mu$, which is usually considered to be very small
(of the order of the Planck scale). However, our intuition in the
present case is that the propagation of the GWs may capture some insights
about these modifications despite the small values of the polymer
scales.

After deriving a tine-dependent effective polymer Hamiltonian using
a novel approach, we derived both nonperturbative and perturbative
analytical expression for the solutions and analyzed them to obtain
further insight into the behavior of such waves. As a result, we found
the following: 
\begin{enumerate}
\item[i)] The form of the waves is modified. More precisely, there is a phase
shift with respect to the classical case. Furthermore, small-amplitude
harmonics are excited. 
\item[ii)] The speed of the waves turns out to be smaller than the speed of
light by a factor $k^{2}\left(\frac{\mathcal{E}_{I}\mu}{4}\right)^{2}$.
This factor not only depends on the polymer scale $\mu$ and the initial
momentum of the perturbations $\mathcal{E}_{I}$, but also on the
wave vector $k$ or, equivalently, the frequency of the waves. Thus,
the higher-energy waves show a greater deviation from the classical
behavior compared to the low-energy waves. 
\item[iii)] The modifications to the classical behavior due to quantum effects
become increasingly visible as the waves travel: the corrections result
in an effective phase shift, which can become of order unity when
$\mathcal{E}_{I}\mu^{2}k^{3}D_{s}$ is of order unity for a distance
$D_{s}$ traveled. 
\end{enumerate}
In a future work, we will proceed to apply our method to the case
where both the background spacetime and the perturbations are effective.

\bibliographystyle{ws-procs961x669}
\bibliography{mainbib}

\end{document}